\documentclass[3p,preprint,hyperref]{elsarticle}
\input{prolegomenaHCT}

\begin{document}

\begin{frontmatter}
\title{Field reparametrization in effective field theories}

\author[torino]{Giampiero Passarino\fnref{support}}
\ead{giampiero@to.infn.it}

\address[torino]{\csumb}

\fntext[support]{\support}

\begin{abstract}
Debate topic for Effective Field Theory (EFT) is the choice of a ``basis'' for $\mrdim = 6$ 
operators\footnote[2]{
{\it{Besides it is an error to believe that rigour is the enemy of simplicity. On the 
contrary we find it confirmed by numerous examples that the rigorous method is at the same 
time the simpler and the more easily comprehended. The very effort for rigor forces us to 
find out simpler methods of proof}}. David Hilbert ``Mathematical Problems'', Bulletin of the 
American Mathematical Society (Jul 1902), 8, 44}. Clearly all bases are equivalent as long 
as they are a ``basis'', containing a minimal set of operators after the use of equations of 
motion and respecting the $SU(3)\,\times\,SU(2)\,\times\,U(1)$ gauge invariance. From a more 
formal point of view a basis is characterized by its closure with respect to renormalization. 
Equivalence of bases should always be understood as a statement for the $\mrS\,$-matrix and not 
for the Lagrangian, as dictated by the equivalence theorem. Any phenomenological approach that 
misses one of these ingredients is still acceptable for a preliminar analysis, as long as it 
does not pretend to be an EFT. Here we revisit the equivalence theorem and its consequences for 
EFT when two sets of higher dimensional operators are connected by a set of non-linear, 
noninvariant, field reparametrizations. 
\end{abstract}
\begin{keyword}
Higgs physics; Standard Model Effective Field Theory
\PACS 12.60.-i \sep 11.10.-z \sep14.80.Bn
\end{keyword}

\end{frontmatter}

\section{Introduction}
The construction of the Standard Model EFT (SMEFT) is based on the fact that experiments occur 
at finite energy and ``measure'' an effective action $\mrS^{\mathrm{eff}}(\Lambda)$;
therefore
\bei

\item[\dnuma] whatever QFT should give low energy 
$\mrS^{\mathrm{eff}}(\Lambda)\,,\;\forall\,\Lambda < \infty$;

\item[\dnumb] one also assumes that there is no fundamental scale above which 
$\mrS^{\mathrm{eff}(\Lambda)}$ is not defined~\cite{Costello2011} and 

\item[\dnumc] $\mrS^{\mathrm{eff}}(\Lambda)$ loses its predictive power 
if a process at $E = \Lambda$ requires $\infty$ renormalized parameters~\cite{Preskill:1990fr}.

\eei
A question that is often raised concerns the ``optimal'' parametrization of the $\mrdim = 6$
basis; once again, all sets of gauge invariant, dimension $d$ operators, none of which is 
redundant, form a basis and all bases are equivalent. For a formal definition of redundancy 
see Sect.~3 of \Bref{Einhorn:2013kja}.

There is no principal obstacle in ``extracting'' (Wilson) coefficients as defined in a particular 
basis. However, certain linear combinations of Wilson coefficients in one basis become a 
single Wilson coefficient in another basis and a mapping of this type that put coefficients 
and (pseudo-)observables in a one-to-one correspondence may seem more appropriate when 
considering LO constraints from electroweak precision data (EWPD).
Even at this level one should be careful since Wilson coefficients mix under renormalization. 
We will analyze under which conditions the bijection can be realized.
Indeed, there is an important lesson to be remembered: it is impossible to redefine the fields 
in such a way that the Lagrangian is unaltered, this is a privilege of the $\mrS\,$-matrix, as
it has been stressed in Sect. II.2.3 of \Bref{YR4}, at least in its original 
version~\cite{NLOnote,Passarino:2016pzb}. There are 
differences between the Lagrangian and the $\mrS\,$-matrix elements; although the Lagrangian 
plays a central role in any theory, observables are related to $\mrS\,$-matrix elements. The 
correct path is: a) computation of off-shell Green's functions, b) normalization of the sources, 
c) amputation of Green's functions and (finally) d) on-shell limit for external legs.

With this in mind, one has to understand why certain transformations are used/useful.
Field transformations of the form $\upPhi' = \mrZ_{\upPhi}\,\upPhi$ ($\mrZ_{\upPhi}$ being
field independent) are a nice way to derive LSZ~\cite{Lehmann:1957zz} factors for the external 
legs (so-called wave-function normalization factors), as it will be discussed in \eqn{LSZ}. They 
are needed to make sure that residues of propagators are $1$ also in the interacting theory: 
this is what it is meant, for instance, by canonical normalization of kinetic terms in SMEFT.

\begin{remark}
Local and non-local transformations have been used by 't Hooft and Veltman~\cite{'tHooft:1972ue} 
in their proof of renormalizability. Non-local transforms always require the addition of ghost 
loops~\cite{Exa}. For local transformations those loops are zero in dimensional regularization 
(integrals of polynomials).
In particular one cannot take a free theory and make it an interacting one by means of field 
transformations. Equivalently we cannot de-interact (even partially) a Lagrangian, see 
Sect.~10.4 of \Bref{'tHooft:1973pz} for a complete discussion.
In other words, making an interaction term (in the Lagrangian) look like what we want does 
not change the physics of the problem and that term will keep contributing to every process it
was contributing before the transformation; we will discuss a simple example with the help of
Fig.~\ref{FTF7} in Sect.~\ref{SMEFTc}.
\end{remark}
\section{Field reparametrizations \label{FT}}
Up for debate is imposing non-linear transformations on a gauge invariant EFT basis and the
correct interpretation of the Equivalence Theorem 
(ET)~\cite{Chisholm:1961tha,Kamefuchi:1961sb,Kallosh:1972ap}. 
Our notations will be as follows (please note the Pauli metric): the scalar field $\upPhi$ 
(with hypercharge $1/2$) is defined 
by
\[
\upPhi = \frac{1}{\srt}\,\left(
\begin{array}{c}
\PH + 2\,\frac{M}{g} + i\,\Ppz \\
\srt\,i\,\Ppm
\end{array}
\right)
\]
$\PH$ is the custodial singlet in $\lpar 2_{\ssL}\,\otimes\,2_{\ssR}\rpar = 1\,\oplus\,3$.
The VEV is $\mrv = 2\,M/g$. Furthermore, $\Ppz, \Pppm$ are Higgs-Kibble ghosts, sometimes
called Goldstone degrees of freedom.

The typical example that we have in mind is the elimination of $\PH\,(\partial\PH)^2$ 
terms from the Lagrangian, but the question is more general and concerns the interpretation of
interaction terms and the idea of rearranging/eliminating certain contributions from the 
Lagrangian. For instance, 
starting from a given set of operators and look for a choice of convenient conventions that 
should allow to calculate physical observables in a ``more transparent way'' (\eg operators 
orthogonally projected or diagonalization in the space of [Wilson coefficients, SM deviations], 
isolation of $\mrdim = 6$ effects in the interaction Lagrangian \etc). 
The question that we want to answer is: can this be done in general and extended beyond LO? 
What is the price to pay given the fact that the Lagrangian (sic) is not invariant under
field reparametrizations that are not gauge transformations? Is it relevant and/or convenient?

For the time being, we start with a simpler example: consider a Lagrangian
\bq
\mcL = \frac{1}{2}\,\lambda^{-2}\,\phi\,\lpar \Box - m^2 \rpar\,\phi + 
       g\,{\overline{\psi}}\,\psi\,\phi + \mrZ\,J\,\phi + 
       {\overline{\mrK}}\,\psi + {\overline{\psi}}\,\mrK \spc
\eq
where $\phi$ is a scalar field and $\psi$ a spinor field. Furthermore, we have added the source 
terms; $\lambda$ reproduces the effect of higher dimensional operators, \eg in SMEFT when the 
Higgs field is replaced by its VEV or when loop corrections are included. The (properly 
normalized) propagation function for the scalar particle is
\bq
\mrZ\,J\,\frac{\lambda^2}{p^2 + m^2}\,\mrZ\,J \spc
\eq
fixing $\mrZ = \lambda^{-1}$. The net effect on the $\mrS\,$-matrix is that, for each
external $\phi$ line, we have a factor $\lambda$. Alternatively, we can define a new
field, $\phi= \lambda\,\phi'$: the Lagrangian is now
\bq
\mcL' =  \frac{1}{2}\,\phi'\,\lpar \Box - m^2 \rpar\,\phi' + 
         g\,\lambda\,{\overline{\psi}}\,\psi\,\phi' + \mrZ'\,J\,\phi' + \dots 
\label{LSZ}
\eq
so that $\mrZ' = 1$. However the $\mrS\,$-matrix elements has a factor $\lambda$ for each 
external $\phi'\,$-line, \eg due to the coupling $g\,\lambda\,{\overline{\psi}}\,\psi\,\phi'$.
This simple example proves that the field redefinition is a matter of taste, the crucial point 
is in the normalization of the source.

We now return to field dependent transformations. Few preliminar facts:
\bei

\item[\dnuma] The scalar manifold is not flat~\cite{Burgess:2010zq}, therefore one cannot get 
rid of $\upPhi\,(\partial\upPhi)^2$ terms ($\upPhi$ is the Higgs doublet) in a gauge invariant 
way. 

\item[\dnumb] Any non-linear transformation of $\PH$ (the Higgs field) will break the symmetries
             (including any broken symmetry) of the theory. 

\item[\dnumc] What to expect from a spin one Lagrangian without symmetries? Note that here
              we are talking about the Lagrangian, see Sect.~\ref{ETiL} for details.

\eei
\subsection{Equivalence Theorem in the literature \label{ETiL}}
It is worth noting that Arzt, in his comprehensive paper~\cite{Arzt:1993gz} is always very careful 
about invariance, \eg see
\bei

\item[a)] statement at p.~$7$, {\it{The variable shift we have performed respects the symmetries 
of the theory $\dots$ Because of this the new Lagrangian explicitly retains all the symmetries of 
the original}};

\item[b)] last paragraph at p.~$10$ where he is, once again, careful in transforming the 
scalar doublet and not the Higgs field alone.

\eei
Indeed, the Equivalence Theorem has been used to derive the Warsaw 
basis~\cite{Grzadkowski:2010es}, where everything is manifestly gauge invariant, what to do with 
non-invariant terms? It seems really trivial, you break gauge invariance and, to get the right 
result, you have to cure your transformation with something equally non invariant. Cui prodest? 
Nevertheless we will show how things actually work.

One should also be careful in stating that the proof of the ET is that only renormalized theories
can be equivalent and it is rigorous only if in both formulations the theory will not contain any 
divergences~\cite{Kallosh:1972ap}: 
{\it{The basic defects of the known formal ``proofs'' of the equivalence theorem are that in all 
of them one compares and asserts the equivalence of unrenormalized quantities, whereas one 
should compare renormalized quantities. In addition to the physical considerations
there are also mathematical reasons for requiring this; owing to the divergences in the 
Lagrangian field theory the unrenormalized quantities simply do not exist}}.

For instance, dimensional regularization is crucial in proving that ghosts are not needed
for local transformations (integrals of polynomials in momenta are zero).
As pointed out by Kallosh and Tyutin, it remains to be proved that the renormalized charges 
are the same in both theories. The conditions for that are only realized in a gauge
theory.

For completeness we recall the main argument used in \Bref{Grzadkowski:2010es}: given a theory 
with a gauge invariant Lagrangian $\Lag(\upPhi\,\,\upPhi^{\dagger})$ (\eg the SM) consider an 
effective Lagrangian 
\bq
\Lag^{(2)}_{\eff} = \Lag + a\,\Ope + a'\,\Ope' \spc
\qquad
\Ope' - \Ope = \mrF\,\frac{\delta\Lag}{\delta\upPhi^{\dagger}} \spc
\label{opeopep}
\eq
where $\mrF$ is some local functional of $\upPhi\,,\,\upPhi^{\dagger}$ and their covariant
derivatives, $\Ope$ and $\Ope'$ are gauge invariant, higher dimensional, operators and $a, a'$ 
are appropriate powers of $1/\Lambda$. The functional $\mrF$ must transform under gauge 
transformations such that the r.h.s. of the (second) \eqn{opeopep} is invariant.
The effect of $\Ope'$ on the $\mrS\,$-matrix corresponding to 
$\Lag^{(1)}_{\eff} = \Lag + a\,\Ope$ is to shift 
$\upPhi^{\dagger} \to \upPhi^{\dagger} + a'\,\mrF$ 
(if the field is real replace with $\upPhi \to \upPhi + a'\,\mrF$ ) at $\mathcal{O}(a')$, \ie we 
can use
\bq
\Lag^{(2)}_{\eff} = \Lag\lpar \upPhi\,,\,\upPhi^{\dagger} + a'\,\mrF \rpar + (a + a')\,\Ope \spp
\label{shift}
\eq
The difference between the two Lagrangians is $\mrF$ times the classical equation of motion 
(EoM) for $\upPhi^{\dagger}$ and the shift $\upPhi^{\dagger} \to \upPhi^{\dagger} + a'\,\mrF$ 
respects the symmetries of the theory~\cite{Arzt:1993gz}, \ie $\upPhi^{\dagger} + a'\,\mrF$ 
transforms as $\upPhi^{\dagger}$. Therefore, thanks to the ET (which in this case is trivial), 
we have equivalence of the $\mrS\,$-matrices and gauge invariant Lagrangians, both the original 
and the shifted one. By repeating the shift the process is continued to all orders in $1/\Lambda$.
The advantage, \ie reducing the number of operators while preserving gauge invariance, 
is self-evident. More details will be given in \appendx{MoEoM}.

Even in this case we could do without the elimination of redundant operators. Indeed, it has 
been pointed out in \Bref{Wudka:1994ny} that, even if the $\mrS\,$-matrix elements cannot 
distinguish between two equivalent operators $\Ope$ and $\Ope'$, there is a large quantitative 
difference whether the underlying theory can generate $\Ope'$ or not. It is equally reasonable 
not to eliminate redundant operators and, eventually, exploit redundancy to check $mrS\,$-matrix 
elements, see \appendx{MoEoM} for details.
\subsection{Equivalence theorem and field reparametrization}
Actually it is not clear, at all, why one needs to perform non-linear $\PH$ transformations
($\PH$ id not a scalar under the group), or any non-linear, non gauge invariant, transformation 
in the SMEFT (once EoMs have been used); in any case there is no added value. There are three  
criteria: wrong, irrelevant and relevant. Let us perform a field transformation on a given
set of higher dimensional operators and assume that it is formally correct:
\bei

\item[i)] the ET tells you that the $\mrS\,$-matrices are equivalent, ergo the
          transformation is irrelevant, as long as propagators have residue one at their 
          poles. If one wants, it is even possible to kill the $\PH\PAQb\PQb$ interaction
          term in the Lagrangian (through a non-local transformation, see \eqn{nonlocal}), does 
          that mean that we get read of the decay? Obviously not.

\item[ii)] If there are stringent arguments to perform the transformation, be aware of
           the balance between gain and complexity.

\eei
On the fact that restricting to a special gauge and using non-linear transformations
makes it extremely hard to go to next-to-leading order (NLO), we hope that there is no 
discussion (otherwise see below).
\begin{example}[Gauge dependence]
Given that it is impossible to get rid of terms with derivatives in some arbitrary gauge, 
the situation is as follows: somebody goes to the unitary gauge, transform $\PH$ with a 
transformation so to get rid of $\PH\,(\PH^2)\,(\partial\PH)^2$.   
Somebody else, wanting to play the same (gauge dependent) game, goes to the $\mrR_{\xi}$ 
gauge and gets rid of cubic and quartic terms with, at least, one $\PH$ derivative. 
This requires a cubic transformations involving $\PH$ and $\Ppz$. How to compare?
Working in the unitary gauge is not as simple as setting the Higgs-Kibble ghosts ($\Ppz, \Pppm$) 
to zero in the Lagrangian, especially when loops are included. It is not our goal to re-describe 
the problem, see Sect.~3 of \Bref{Exa} in order to re-discover the issue.
\end{example}
By the time we accept a Lagrangian which is not invariant a crucial test is given by off-shell 
Ward-Slavnov-Taylor (WST) identities (which means unitarity) or BRST invariance. Even more, an 
unstable particle is characterized by its complex pole which requires its  self-energy computed
off-shell. In that case one has to prove that the latter does not depend on the transformation. 
Let us forget, for a moment, gauge invariance of the Lagrangian and blindly perform the 
transformation.
\begin{example}[Question time]

\bei

\item[\dnumA] Are the two Lagrangians (original and transformed) the same? No, they are
              different;

\item[\dnumB] what about off-shell amplitudes, are they the same? No, not even for a gauge 
              invariant transformation.

\eei
\end{example}
\begin{remark}
To do a correct job one has to take into account the transformed Lagrangian, the Jacobian and 
the new couplings to the sources (it is the LSZ formalism). To study the effect 
on mass-shell one needs to renormalize Green's functions, \ie off-shell Green's functions 
are changed.
\end{remark}
\subsection{Scattering matrix vs. Lagrangian}
What about the $\mrS\,$-matrix? To make them equivalent, accepting that the assumption of the
theorem are not violated, would require a very complex set of operations that involve 
the correct treatment of sources (see below). Extending the calculation for any process 
beyond LO is remarkably difficult. 

It is worth noting that in QCD scaleless bubbles are zero, that is why most people tend to 
forget LSZ. The core of the exercise is to work out the transformed theory within the framework of
the original one and to look at the effect of the ``new'' vertices.
To be precise, diagrams can be partitioned into classes where there is complete cancellation 
after the transformation (use that the Klein-Gordon operator is minus the inverse propagator) but 
the treatment of sources is crucial for the rest. Therefore, always assuming that it works, one 
should transmit a complete information for any new ``basis'' proposal, both the Lagrangians and 
the field transformations. 

The use of EoM in constructing the ``Warsaw'' basis leaves the Lagrangian invariant, the ET 
works in the same way in all gauges and we do not need to bother about details of the 
transformation, the Lagrangian without redundant operators is the ``starting point''. A 
subsequent, non invariant, field reparametrization is a different story: with the (transformed) 
Lagrangian alone one will never be able to perform a complete calculation, the transformation 
is needed in order to evaluate the Jacobian (trivial only for a local transformation) and the 
gauge dependent change in the source factors, \ie the shift in wave-function renormalization. 
As a matter of fact, even the ``starting point'' is not free from arbitrariness in the choice
of gauge invariant operators. A classification scheme has been proposed in \Bref{Einhorn:2013kja}
where it is shown that the ``Warsaw'' basis satisfies the criterion dictated by the scheme.

The general statement is that a given effect could be modeled by several different combinations of 
operators at a fixed order in the SMEFT, depending on the basis. 
This statement, once again, should always be understood as referring to physical observables,
\ie to the $\mrS\,$-matrix.
Existing tools are often bound to a given basis choice; any platform designed for supporting 
the reuse of results derived in the context of one (reparametrized) basis in another 
(original) basis should take this trivial fact into account.
Is it optimal to perform field redefinitions? No. Let us clarify this point: for the sake
of simplicity we start with a non-gauge theory,
\bq
W[J] = \int \Bigl[ D\phi \Bigr]\,\exp\,\{ \mrS(\phi) + \int d^4x\,J\,\phi\} \spc
\label{Wdef}
\eq
where $\mrS(\phi)$ is the action. Obviously we may change the integration variable without 
changing the integral, $\phi = \psi + \mrF(\psi)$; it is crucial that $\mrF$ starts at $\psi^2$.
Note that $\phi= a\,\phi$ is standard renormalization, \ie. LSZ or wave-function renormalization.
We obtain
\bq
W[J] = \int \Bigl[ D\psi \Bigr]\,\mathrm{det}\lpar 1 + \mrF' \rpar\,
\exp\,\{ \mrS(\psi + \mrF) + \int d^4x\,J\,(\psi + \mrF) \} \spc
\label{eqone}
\eq
where $\mrF' = \delta \mrF/\delta \phi$. If we have started with and action $\mrS(\psi + \mrF)$ 
we would have derived
\bq
W'[J] = \int \Bigl[ D\psi \Bigr]\,\mathrm{det}\lpar 1 + \mrF' \rpar\,
\exp\,\{ \mrS(\psi + \mrF) + \int d^4x\,J\,\psi \} \spc
\eq   
and $W$ is different from $W'$ (of course $W[0] = W'[0]$), not to mention 
\bq
W''[J] = \int \Bigl[ D\psi \Bigr]\,
\exp\,\{ \mrS(\psi + \mrF) + \int d^4x\,J\,\psi \}  \spp
\eq
\begin{remark}
What do we do if we have an action for which we do not know that it is a somehow transformed 
other action? A Lagrangian does not unambiguously define correlators. Questionable if you take a 
non invariant action that is the transformation of an invariant one. 
\end{remark}
Requiring sources to be physical will not help since there is no restriction on the 
$\PH$ source in the SM and in the SMEFT, see Eq.~(4.1) {-} Eq.~(4.2) of \Bref{Exa}.
Of course, for a complete discussion, see Sect.~4 of \Bref{Kallosh:1972ap}, in
particular the discussion on wave-function normalization constants in different gauges.
As explicitly shown in \Bref{'tHooft:1972ue} the proof of the equivalence theorem is also the 
proof of unitarity.

\begin{figure}[ht]
\vspace{-3.cm}
\centering
\includegraphics[width=0.9\textwidth, trim = 30 250 50 80, clip=true]{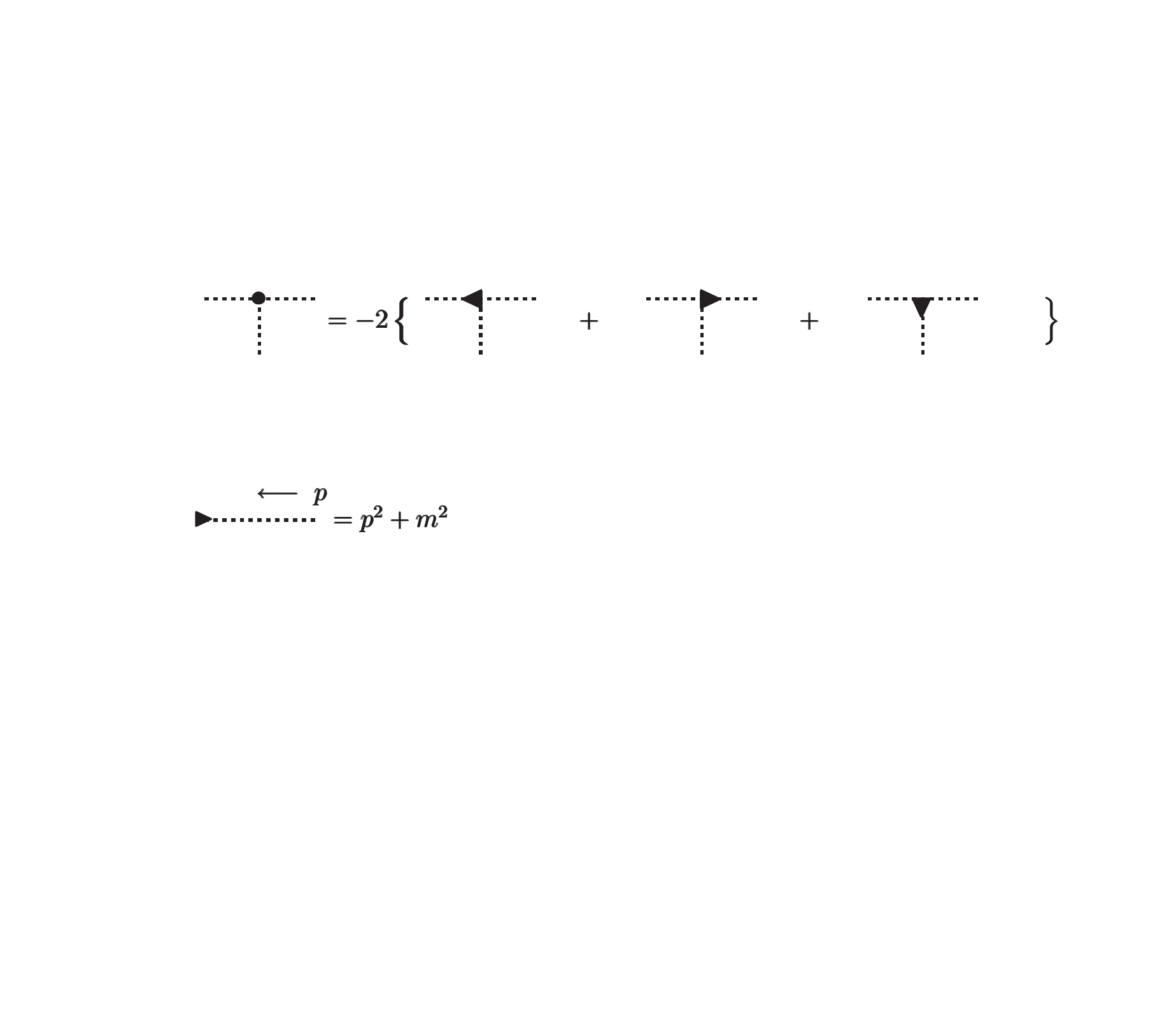}
\vspace{-6.cm}
 \caption{Special vertices in a non-linear field transformation.}
\label{FTF1}
\end{figure}

\begin{figure}[ht]
\vspace{-2.cm}
\centering
\includegraphics[width=0.8\textwidth, trim = 30 250 50 80, clip=true]{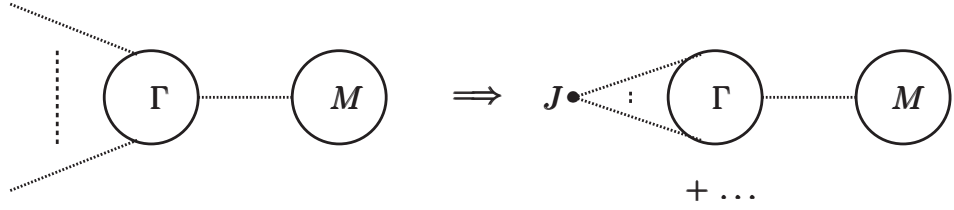}
\vspace{-6.cm}
 \caption{The vertex $\Gamma$ needed to construct $\Delta\mrZ$ in \eqn{Zfact}. $M$ is an 
amputated Green's function, the lines to the left of $\Gamma$ correspond to transformed 
fields $\psi$, the one connecting $\Gamma$ with $M$ corresponds to the original field $\phi$.}
\label{FTF2}
\end{figure}

Note that, adding to the functional integral approach, we can provide a purely diagrammatic proof; 
indeed all manipulations carried out with functional integrals can be explicitly checked using 
the diagram technique, as discussed in \Brefs{Gervais:1976ws,Salomonson:1976yn,Arzt:1993gz} where 
the special role of dimensional regularization is emphasized.  
The diagrammatic proof is based on the following identity:
consider the transformation $\phi \to \psi + a\,\psi^2$ and its effect on the part of the
Lagrangian quadratic in $\phi$,
\bq
\frac{1}{2}\,\phi\,\lpar \Box - m^2 \rpar\,\phi \spp
\label{quad}
\eq
We obtain two ``special'' vertices, with three and four $\psi$ fields, that can be decomposed 
according to the rule of Fig.~\ref{FTF1}.
and the propagator is $1/(p^2 + m^2)$; special vertices will be denoted by a black triangle.
Of course, there will be additional new vertices coming from $\Lag_{\inte}(\psi)$, to be denoted
by a black bullet. The rule is easily generalizable to arbitrary polynomial transformations. For 
instance, in SMEFT, we will have $\PH \to \PH + a\,\PH^n$ or $\PH \to \PH + \mrP(\PH\,,\,\dots)$ 
where $\mrP$ is a polynomial in the fields and in their derivatives. Note that the special 
vertices on the r.h.s. of Fig.~\ref{FTF1} do not contribute to the $\mrS\,$-matrix if the line 
is external (no one-particle pole): this is how one should understand elimination of
$\mrdim = 6$ contributions to propagators through EoM. Consequences will be relevant when
trying to isolate effects in the interaction, as we will show with the example of \eqn{exex}
and of Fig.~\ref{FTF7}. 
\begin{remark}[Shortly:] 
a black triangle pointing towards a line cancels the corresponding propagator (shrinking the
line to a point) and changes sign of the diagram, adding the proper combinatorial factor,
see \appendx{cf}.
\end{remark}
Note that, when moving to SMEFT, the transformation $\PH \to \PH + \mrP$ is not a gauge 
transformation, therefore we loose the group property which means that finite transformations 
($a$) do not follow as a consequence of infinitesimal ones ($\delta\,a$).
\begin{proposition}
The renormalization constants of the two theories, say $\Lag(\phi)$ and $\Lag(\psi + \mrF(\psi))$, 
are related by 
\bq
\mrZ_{\phi} = \mrZ_{\psi}\,\lpar 1 + \Delta\mrZ \rpar^2 \spc
\label{Zfact}
\eq
where $\mrZ_{\phi}$ and $\mrZ_{\psi}$ are such that the $\phi$, $\psi$ propagators have residue 
one at their poles, \ie one has to be careful and write $\mrZ^{-1/2}_{\phi}\,J\,\phi$ \etc in 
the equations above.
At the same time amputated Green's functions satisfy
\bq
\mrG^{(k)}_{\phi} = \lpar 1 + \Delta\mrZ \rpar^k\,\mrG^{(k)}_{\psi} \spc
\label{Gfact}
\eq
which is the main result of \Bref{Kallosh:1972ap}, proving the theorem.
\end{proposition}
The $\Delta\mrZ$ factor, as discussed in \Bref{Kallosh:1972ap}, must be derived in terms of a 
vertex $\Gamma$ computed on the mass shell of the $\psi$ line connecting $\Gamma$ with the 
amputated Green's function $M$, see Fig.~\ref{FTF2}. Here we have shown only one of the 
terms that contribute, most terms in the list give zero result because they do not exhibit 
the one-particle pole, \eg a source emitting $n$ lines directly connected to $M$.
Therefore, it is formally required to start with
\bq       
W[J] = \int \Bigl[ D\psi \Bigr]\,\mathrm{det}\,
       \lpar 1 + \mrF' \rpar\,
       \exp\{ \mrS(\psi + \mrF) + \int d^4x\, (\mrZ_{\psi})^{-1/2}\,J\,\psi \} \spp
\label{eqtwo}
\eq  
Admittedly, one can neglect the Jacobian if $\mrF$ is local. It is worth noting that \eqn{eqone} 
defines the theory in the $\phi$ framework while \eqn{eqtwo} defines the theory in the 
$\psi$ framework. 

\begin{figure}[ht]
\vspace{-1.cm}
\centering
\includegraphics[width=0.8\textwidth, trim = 30 250 50 80, clip=true]{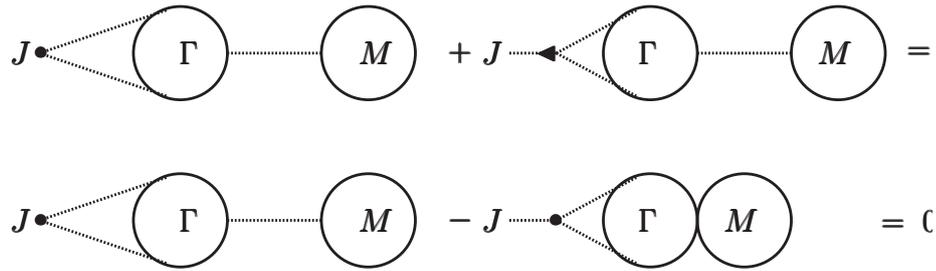}
\vspace{-6.cm}
 \caption{Example of a class of diagrams (exhibiting one-particle poles) where cancellation occurs.
The first diagram is part of the change in the source normalization $\mrZ$ while the second
gives a change in the Green's function.}
\label{FTF6}
\end{figure}

Perhaps it is easier to understand the theorem at the diagrammatic level, without having to
split the new diagrams into a contribution to the Green's functions or to the source 
normalization. The ET is based on the fact that diagrams can be allocated to classes within 
which they cancel, as shown in Fig.~\ref{FTF6}.
\subsubsection{The SMEFT case \label{SMEFTc}}
\begin{figure}[ht]
\vspace{-2.cm}
\centering
\includegraphics[width=0.9\textwidth, trim = 30 250 50 80, clip=true]{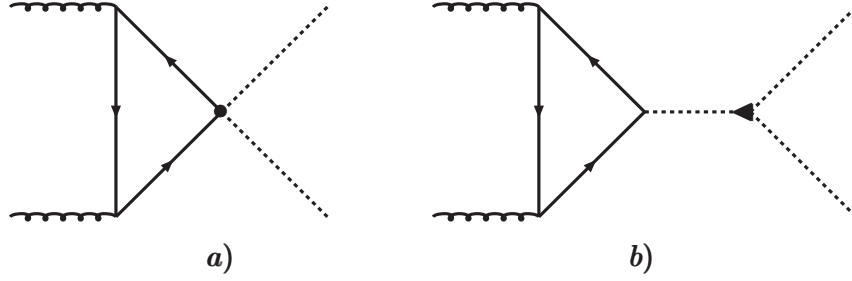}
\vspace{-7.cm}
 \caption{Diagrams contributing to $\Pg\Pg \to \PH\PH$ after the non-linear transformation,
Special vertices are denoted by $\bullet$ and $\blacktriangleleft$. The diagram a) is needed
to cancel diagram b).}
\label{FTF3}
\end{figure}

Generally speaking, one will have to produce a complete account for any pair of bases that
are related by a transformation; once again, the Lagrangian alone is not enough. When one takes 
into account source normalization (\eg $\mrZ_{\phi}$ and $\mrZ_{\psi}$) it will be easy to 
discover that attempting a field redefinition actually makes the whole calculation 
much, much more involved. Who wants to compute $\Gamma$ at higher orders? Furthermore, the 
derivation is completely screwed up if one does not take into account that, now, the source can 
emit/absorb multi particles (non-linear transformation), something that is not seen from the 
Lagrangian point of view. We now consider few specific examples. 
\begin{example}[Off-shell]
Most of the analyses for off-shell Higgs physics are done (possibly using two different
codes) by using off-shell production$\,\times\,$propagator$\,\times\,$off-shell decay. After the 
transformation the latter, containing diagram b) of Fig.~\ref{FTF3}, is not gauge invariant, even 
at LO; in the SM one must go to one loop before seeing the effect of off-shell gauge parameter 
dependence. See the $\Pp\Pp \to \PAQt\PQt\PAQb\PQb$ or the $\Pp\Pp \to \PH\PH$ examples, \eg 
in $\Pg\Pg \to \PH\PH$, as illustrated in Fig.~\ref{FTF3} where the two diagrams cancel.
In the production$\,\times\,$decay approach it would be wrong to neglect diagram b) using
EoM.
\end{example}
How to deal with higher orders and renormalization is, yet, another problem.
Consider the transformation $\phi \to \psi + \mrF(\psi)$, given
\bq
< \phi(x)\,\phi(y) > = < \psi(x)\,\psi(y) >_o + < \psi(x)\,\mrF(\psi(y)) >_o + \dots
\eq
one has to prove that the position of the pole of the $\phi$ propagator coincides with
the position of the pole of the $\psi$ propagator (in a spontaneously broken theory).
The index ``o'' in the r.h.s. denotes that the corresponding quantities are computed
in the original theory. We will only indicate the main idea, with the essential details,
\begin{example}[Tadpoles]
Let us consider tadpoles. Taking the Higgs potential and performing the transformation
$\PH \to \PH + g\,a/M\,\PH^2$ and using $P = \upPhi_{\mathrm{pot}} + 1/2\,M^2_{\PH}\,\PH^2$
(where $\upPhi_{\mathrm{pot}}$ is the scalar potential) we obtain
\bq
P \to P - a\,\Bigl[  2\,\beta_{\PH}\,\PH^2 + \frac{g}{M}\,\beta_{\PH}\,\PH^3 +
      \frac{3}{4}\,g^2\,\frac{M^2_{\PH}}{M^2}\,\PH^4 + 
      \frac{1}{8}\,g^3\,\frac{M^2_{\PH}}{M^3}\,\PH^5 \Bigr] + \mathcal{O}(a^2) \spc
\eq
where $\beta_{\PH}$ is fixed order-by-order in perturbation theory to cancel tadpoles. Therefore, 
neglecting all other fields, at one loop we derive 
\bq
\beta_{\PH}= - \frac{3}{64}\,\frac{g^2}{\pi^2}\,\frac{M^2_{\PH}}{M^2}\,\mrA_0\,
             \lpar 1 + 4\,a \rpar \spc
\eq
where $\mrA_0$ is the one point function of argument $M_{\PH}$. This is what we obtain by looking
at the Lagrangian, \ie. by including only part of the diagrams. For instance we included
the diagrams of Fig.~\ref{FTF4} containing a transformed vertex.
\begin{figure}[ht]
\vspace{-3.cm}
\centering
\includegraphics[width=0.9\textwidth, trim = 30 250 50 80, clip=true]{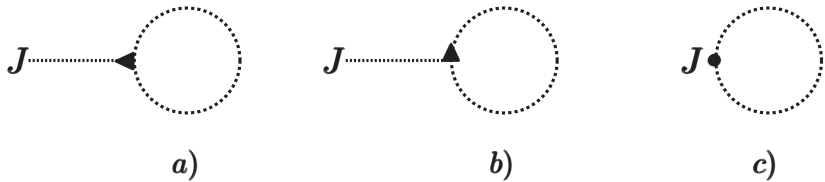}
\vspace{-7.cm}
 \caption{Tadpoles after the non-linear transformation. Diagram b) is zero in dimensional
regularization while diagrams a) and c) cancel.}
\label{FTF4}
\end{figure}
Obviously, the correct treatment of $\beta_{\PH}$ makes the whole procedure unnecessarily complex.
\end{example}
\begin{example}[Self-energy]
The result is also ``a'' dependent when we carelessly compute the one loop $\PH$ 
self-energy ($\Sigma_{\PH}$), \eg by including diagram a) and not diagram b) of Fig.~\ref{FTF5},
plus other, less intuitive, cancellations. 
\begin{figure}[ht]
\vspace{-3.cm}
\centering
\includegraphics[width=0.9\textwidth, trim = 30 250 50 80, clip=true]{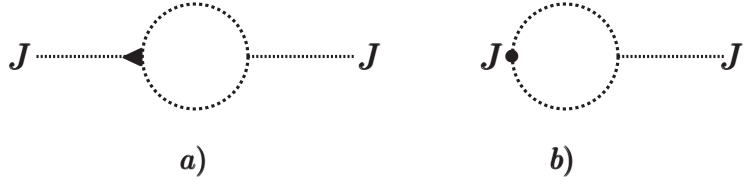}
\vspace{-7.cm}
 \caption{Example of cancellation in the transformed Higgs self-energy.}
\label{FTF5}
\end{figure}
Note that $\Sigma_{\PH}$ is needed for renormalization, \ie for fixing counterterms. Clearly, 
the road to a unique, renormalized, $\mrS\,$-matrix is not simple. N.B. after the transformation, 
if we perform a generic gauge transformation, there will be tadpoles depending of the parameters 
of the gauge transformation.
\end{example}
\subsection{De-interacting/rearranging the Lagrangian? \label{deint}}
\begin{figure}[ht]
\vspace{-2.cm}
\centering
\includegraphics[width=0.9\textwidth, trim = 30 250 50 80, clip=true]{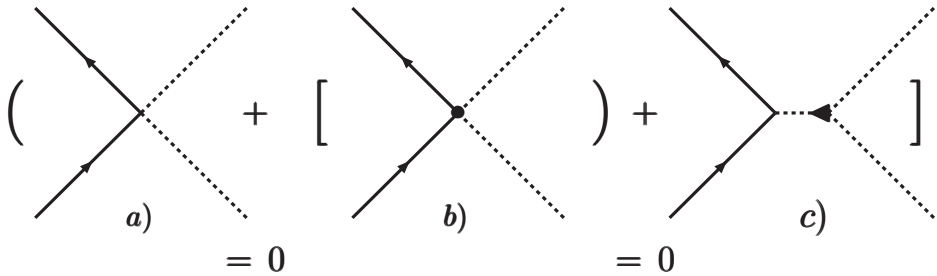}
\vspace{-7.cm}
 \caption{Cancellation, non-cancellation of $\PAQt\PQt\PH^2$. The effect of diagram c) is to
enforce equivalence at the $\mrS\,$-matrix level. Diagrams with special vertices 
$\blacktriangleright$ on the external legs are not included since they do not contribute to 
the $\mrS\,$-matrix. Therefore, b) + c) = $0$.}
\label{FTF7}
\end{figure}

In this Section we give more details on non-invariant field reparametrizations aimed to 
modify/rearrange terms in the interaction Lagrangian. Suppose that we want to transform $\PH$ 
so that the $\PAQt\PQt\PH^2$ term, generated by $\Ope_{\PQt\,\upphi}$ (see Tab.~2 of 
\Bref{Grzadkowski:2010es}), is eliminated. At $\mcO(\Lambda^{-2})$ the transformation will be
\bq
\PH \to \PH + a\,\PH^2 \spc
\quad
a = - \frac{3}{4}\,\frac{g}{\myGF\,\Lambda^2\,M}\,\atp \spp
\label{exex}
\eq
We can describe the situation with the help of Fig.~\ref{FTF7}; there is the original diagram a) 
and the is diagram b) containing a new vertex ($\bullet$) originating from the interaction 
Lagrangian after the transformation of \eqn{exex}. The two cancel by construction. However from 
the Lagrangian quadratic in $\PH$, \eqn{quad}, we generate diagram c) with a special 
($\blacktriangleleft$) vertex (introduced in Fig.~\ref{FTF1}).
Cancellation can be seen both ways, ET is based on the fact that b) and c) cancel, therefore
the effect of $\atp$ in $\PAQt\PQt \to \PH\PH$ will survive. 
\begin{remark}
This is precisely the meaning of {\it{the $\mrS\,$-matrix does not change}}. For the same 
reasoning the transformation will not change the $\mrS\,$-matrix elements for 
$\PAQt\PQt \to \PH\PH\PH$ and $\PAQt\PQt \to \PH\PH\PH\PH$ will remain zero at tree level; after 
all, $\PH\,\Box\,\PH^2$ (although generated by $\Ope_{\upphi\,\scriptscriptstyle{\Box}}$), is not 
a gauge invariant operator and, more important, cannot be completely eliminated through the EoM.
Therefore, if one wants to swap interaction terms this is done at the price of working with
a non-invariant Lagrangian even though the $\mrS\,$-matrix is the same. The price to pay is
a non trivial $\mrZ$ factor. Additional consideration will ge given in \appendx{MoNIR}.
\end{remark}
\begin{example}[Non local transformations]
For completeness we will show that any term of the form $\PH\,\mrF$ where $\mrF$ may depend
on all the SM fields can always be ``cancelled'' from the Lagrangian using a non local
transformation,
\bq
\PH(x) \to \PH(x) + i\,\int\,d^4y\,\Delta(x - y)\,\mrF(y) \spc
\quad 
\Delta(z)= \frac{1}{(2\,\pi)^4\,i}\,\int\,d^4p\,\frac{\exp\{i\,\spro{p}{z}\}}{p^2 + M^2_{\PH}} \spp
\label{nonlocal}
\eq
It is enough to observe that $(\Box_x - M^2_{\PH})\,\Delta(x - y) = i\,\delta^{(4)}(x - y)$. 
The transformation of \eqn{nonlocal} allows us to cancel (in the Lagrangian) all terms,
not only those of the form $\upPhi^{\dagger}\,\upPhi$ times a $\mrdim = 4$ operator.
In the non local case however, ghosts are required to show equivalence of the $\mrS\,$-matrices.
\end{example}
%
%
%
\subsection{Additional remark} 
The source $J$ in \eqn{eqone} behaves by definition as a scalar under reparametrizations of the
corresponding field $\phi$. In particular the term $J\,\phi$ is not a scalar under
reparametrizations $\phi = \psi + \mrF(\psi)$, it is therefore impossible to make $W[J]$
defined in \eqn{Wdef} and the effective action, $\Gamma[\Phi]$ of \eqn{effA}, 
reparametrization-invariant quantities simultaneously.
\bq
\exp\{\Gamma[\Phi]\} = 
\int \Bigl[ D\phi \Bigr]\,\exp\,\{ \mrS(\phi) + \int d^4x\,J\,(\phi - \Phi)\} \spc
\label{effA}
\eq
To go deeper into the subject we observe that Vilkovisky~\cite{Vilkovisky:1984st} has argued 
as follows: by requiring additionally that the effective action be invariant under local 
invertible changes in the choice of basic field variables, one can construct a natural unique 
gauge-invariant effective action, \ie $1$PI Green's functions are invariant off-shell 
(in Vilkovisky approach). In other words, a reparametrization-invariant effective action requires 
the so-called logarithmic map~\cite{Vilkovisky:1984st,Batalin:2012rv}.

A more formal study of field diffeomorphisms for free and interacting quantum fields has been
performed in \Bref{Kreimer:2016jxo} with the result that the theory is invariant if and only if 
kinematic renormalization schemes are used.

Finally, what happens if we require reparametrization invariance of the SMEFT Lagrangian?
As observed in \Bref{Finkemeier:1997re,Luke:1992cs}, in the context of heavy quark effective
theory, this leads to severe constraints for the couplings in the effective Lagrangian; in any 
case, neither the form of the Lagrangian nor the form of the reparametrization transformation 
is unique.
\section{Conclusions}
In this paper we have examined the relation between two sets of higher dimensional operators
that are connected by non-linear, gauge dependent, field reparametrizations. SMEFT is a
quantum construct and there is no fundamental scale and no order in perturbation theory above 
which SMEFT is not defined~\cite{Costello2011}. 
That SMEFT loses its predictive power at some $E = \Lambda$ requiring an infinite number of
renormalized parameters is another story; we are not focusing on the NLO numerical impact
but on the internal consistency of the theory. 

The role of the Equivalence Theorem is often underestimated or even misunderstood, oversimplifying 
a complex situation; we have illustrated the steps that are needed in order to reproduce
the correct $\mrS\,$-matrix, a non trivial exercise when gauge dependent reparametrizations are 
involved. In particular the role of wave-function normalization becomes critical. 
It is highly desirable that any theory of SM deviations presents its predictions in terms of
gauge invariant (pseudo-)observables and not in terms of interaction Lagrangians.  
There is a huge difference between starting from a phenomenological (limited) set of
higher dimensional operators and proving a bijection with a basis.
\section{Acknowledgments}
I gratefully acknowledge several important discussions and a productive collaboration with 
M.~Trott. 
\section{Appendix: The role of the combinatorial factors \label{cf}}
\begin{figure}[ht]
\vspace{-1.cm}
\centering
\includegraphics[width=1.\textwidth]{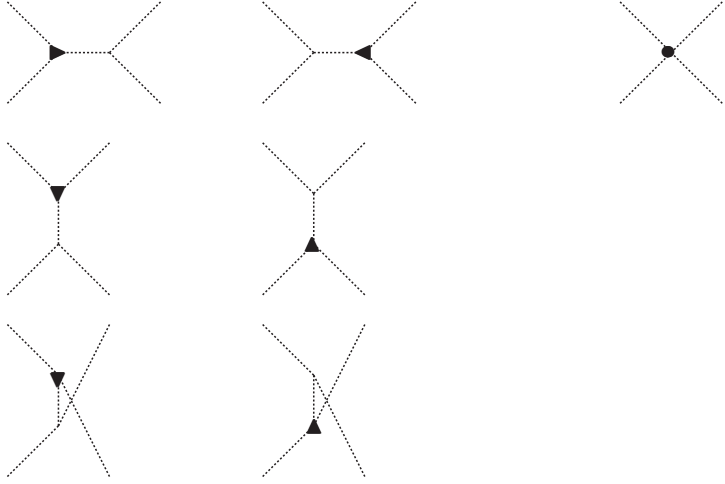}
\vspace{-14.cm}
 \caption{The tree-level four point function in the theory described by the Lagrangian of 
\eqn{lagex}. The diagrams shown contain vertices generate after the transformation 
$\phi \to \phi + a\,\phi^2$}.
\label{FTF8}
\end{figure}

In this Appendix we will give more details about the diagrammatic proof of the ET. Consider
a Lagrangian
\bq
\Lag= \frac{1}{2}\,\phi\,\lpar \Box - m^2 \rpar\,\phi + \lambda\,\phi^3 \spp
\label{lagex}
\eq
There is only one vertex with $3$ lines and equal to $3\,!\,\times\,\lambda$. Performing the 
transformation $\phi \to \phi + a\,\phi^2$ (up to $\mcO(a)$) generates a vertex with $4$ lines 
and equal to $3\,\times\,4\,!\,\times\,\lambda\,a$. In Fig.~\ref{FTF8} we show the diagrams 
containing ``new'' vertices.
The $6$ diagrams in the left part od Fig.~\ref{FTF8} cancel the internal propagator and give
$6\,\times\,3\,!\,\times\,( - 2\,\lambda\,a)$ while the contact term gives 
$3\,\times\,4\,!\,\times\,\lambda\,a$, \ie they cancel.

\begin{figure}[ht]
\vspace{-1.cm}
\centering
\includegraphics[width=0.9\textwidth]{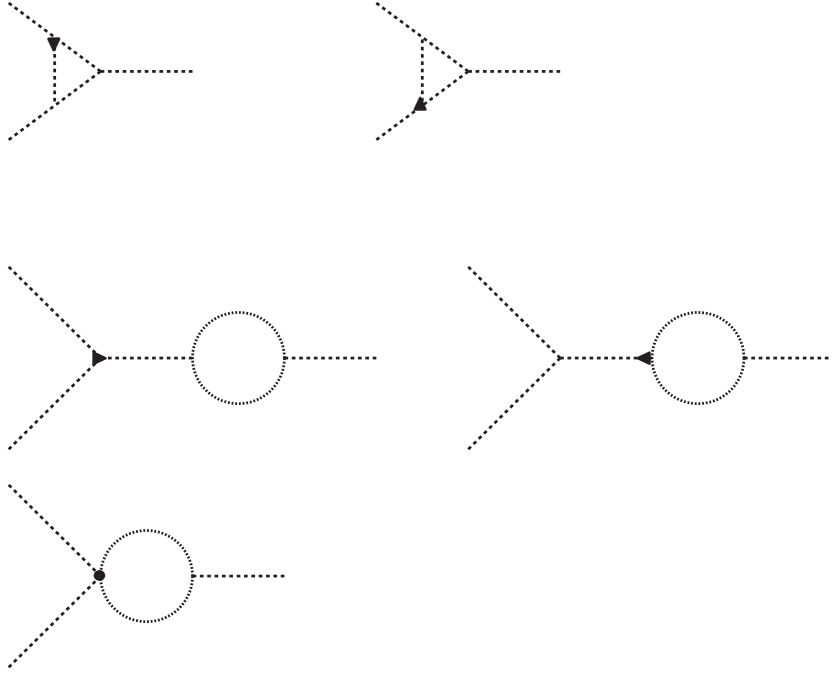}
\vspace{-10.cm}
 \caption{The one-loop three point function in the theory described by the Lagrangian of 
\eqn{lagex}. The diagrams shown contain vertices generate after the transformation 
$\phi \to \phi + a\,\phi^2$}.
\label{FTF9}
\end{figure}

Consider now the one-loop three point function. In Fig.~\ref{FTF9} we show the diagrams 
containing ``new'' vertices. Each diagram in the first row gives 
$(3\,!)^2\,\times\,( - 2\,\lambda^2\,a)$.
Each diagram in the second row gives 
$\frac{1}{2}\,\times\,(3\,!)^2\,\times\,( - 2\,\lambda^2\,a)$, where $1/2$ is
the combinatorial factor. The diagram in the last row gives 
$\frac{1}{2}\,\times\,3\,!\,\times\,4\,!\,\times\,\lambda^2\,a$ where $1/2$ is the combinatorial 
factor. The total is zero.
\section{Appendix: More on EoM \label{MoEoM}}
The content of \eqns{opeopep}{shift} deserves additional comments. In \eqn{shift} we have two
terms, $\Lag\lpar \upPhi\,,\,\upPhi^{\dagger} + a'\,\mrF \rpar$ and $(a + a')\,\Ope$. Does that
mean that we can use $\Lag\lpar \upPhi\,,\,\upPhi^{\dagger} + a'\,\mrF \rpar =
\Lag\lpar \upPhi\,,\,\upPhi^{\dagger}\rpar$? Only a posteriori. To study a realistic example, 
suppose that we enlarge the Warsaw basis by adding a new operator
\bq
\Ope^{(2)}_{\upphi\,\scriptscriptstyle{\PD}} =
\lpar \upPhi^{\dagger}\,\upPhi \rpar\,\Bigl[ \lpar \mrD_{\mu}\,\upPhi \rpar^{\dagger}\,
\mrD^{\mu}\,\upPhi  \Bigr] \spp
\eq
As shown in \Bref{Grzadkowski:2010es} this operator is redundant and is equivalent, through the
EoM, to a linear combination of $\Ope_{\upphi}$, $\Ope_{\upphi\,\scriptscriptstyle{\Box}}$,
$\Ope_{\Pf\,\upphi}$ and $\mrdim = 4$ operators.
The best way of dealing with the $\mrdim = 4$ operators is to perform a shift in the parameters
such that the quadratic part of the Lagrangian, after shifting the fields, coincides with the
$\mrdim = 4$ SM Lagrangian, see \Brefs{Passarino:2012cb,Ghezzi:2015vva}.
If we compute vertices with $3$ fields, one of them being $\PH$, we observe that $\aptD$ can
be absorbed by shifting
\bq
\apBox\to \apBox + \frac{1}{2}\,\aptD \spc \quad
\aup \to \aup - \frac{1}{2}\,\aptD \spc \quad
\adp \to \adp + \frac{1}{2}\,\aptD \spc \quad
\ap \to \ap + \frac{1}{4}\,\frac{M^2_{\PH}}{M^2} \spp
\label{wcshift}
\eq
The request that the functional forms of $\Lag_2(\mathrm{SMEFT})$ and $\Lag_2(\mathrm{SM})$ are
the same (where $\Lag_2$ is the quadratic part of the Lagrangian) implies, among other things, 
the shift
\bq
M_{\PH} \to M_{\PH}\,\Bigl[ 1 + \frac{g_6}{4}\,\Bigl(
          {\overline{\apD}} - 4\,{\overline{\apBox}} + 24\,\frac{M^2}{M^2_{\PH}}\,
          {\overline{\ap}} + 3\,\aptD \Bigr) \Bigr] \spc
\eq
where $g_6= 1/(\myGF\,\Lambda^2)$. Consider now vertices with $4$ fields; as an example, we 
consider the process $\PWpmu(p_1) + \PWmnu(p_2) \to \PH(p_3) + \PH(p_4)$. After shifting the 
Wilson coefficients as in \eqn{wcshift} we have a contact diagram
\bq
\mrV_{\PW\PW\PH\PH} = g^2\,g_6\,\delta_{\mu\nu}\,\frac{1}{4}\,\lpar
             {\overline{\apD}} - 4\,{\overline{\apBox}} + 3\,\aptD \rpar + \dots
\eq
where the shifted parameters of \eqn {wcshift} have benn denoted with a bar. Therefore, there are
terms, proportional to $\aptD$, that survive. This is exactly the role played by
$\Lag\lpar \upPhi\,,\,\upPhi^{\dagger} + a'\,\mrF \rpar$.
The $\PH\PW\PW$ vertex is
\bq
\mrV_{\PH\PW\PW} = - g\,M\,\delta_{\mu\nu} + g^2\,g_6\,\delta_{\mu\nu}\,\frac{1}{4}\,M\,\lpar
                   {\overline{\apD}} - 4\,{\overline{\apBox}} \rpar + \dots
\eq
with a $\PH\PH\PH$ vertex 
\bqa
\mrV_{\PH\PH\PH} &=& - \frac{3}{2}\,g\,\frac{M^2_{\PH}}{M} +
g\,g_6\,\Bigl[
     \frac{9}{8}\,\frac{M^2_{\PH}}{M^2}\,
        \lpar {\overline{\apD}} - 4\,{\overline{\apBox}} \rpar -
       \frac{1}{4}\,\frac{P^2 + M^2_{\PH}}{M}\,
       \lpar {\overline{\apD}} - 4\,{\overline{\apBox}} - 3\,\aptD \rpar 
\nl
{}&-& \frac{1}{4}\,\frac{(p^2_3 + M^2_{\PH}) + (p^2_4 + M^2_{\PH})}{M}\,
       \lpar {\overline{\apD}} - 4\,{\overline{\apBox}} - 3\,\aptD \rpar \Bigr] + \dots
\eqa
where $P = p_1 + p_2$. By adding the contact diagram to the $s\,$-channel diagram
we obtain the total, explicit, $\aptD\,$-dependent part of the process at $\mcO(g^2\,g_6)$,
\bq
- \frac{3}{4}\,g^2\,g_6\,\aptD\,\delta_{\mu\nu}\,\frac{1}{P^2 + M^2_{\PH}}\,
       \Bigl[ \lpar p^2_3 + M^2_{\PH} \rpar + \lpar p^2_4 + M^2_{\PH} \rpar \Bigr] \spc
\eq
which does not contribute to the $\mrS\,$-matrix. Once again, the Lagrangian is explicitly
$\aptD$ dependent even after the shift of \eqn{wcshift}, only the $\mrS\,$-matrix is independent.
In going beyond LO one should always remember that Green's functions are divergent and it is not
until renormalization is performed and $\mrS\,$-matrix elements are computed that we can keep ET 
at bay. Given $\Lag_{\eff}$ of \eqn{shift}, the correct statement is 
\bq
\frac{\partial}{\partial a'}\,\mrS\Bigl[ \Lag_{\eff} \Bigr]\,\bmid_{a+a'= \mathrm{fixed}} = 0 \spc
\eq
where $\mrS[\Lag]$ is the $\mrS\,$-matrix associated to $\Lag$. Note that the whole procedure is
manifestly invariant.
\section{Appendix: More on non invariant reparametrizations \label{MoNIR}}
In this Appendix we elucidate the question of calculating physical observables in 
a ``more transparent way'' by means of field reparametrizations. As we have already seen in
Sect.~\ref{deint} it is possible to swap an interaction term $\PAQt\PQt\PH^2$ in favour
of $\PH^3$ and of $\PH\,\Box\,\PH^2$, at the price of having a non invariant Lagrangian, \ie
we are not dealing with a rescaling of the Wilson coefficient for 
$\Ope_{\upphi\,\scriptscriptstyle{\Box}}$.

When we compute $\PAQt\PQt \to \PH\PH$ (at LO) in the original theory there are $4$ diagrams,
the $s\,$-channel $\PH$ exchange, a contact interaction and $t,u\,$-channel $\PQt$ exchange.
Killing the contact term still leaves $3$ diagrams and the same $\mrS\,$-matrix element. 
Moving NLO requires a careful treatment of the $\PH$ wave function. 

Another example concerns the transformation 
\bq
\PH \to \PH - \frac{1}{8}\,\frac{g\,g_6}{M^2}\,\lpar \apD - 4\,\apBox \rpar\,
              \lpar M\,\PH^2 + \PH^3 \rpar \spc
\label{cubic}
\eq
that changes $3\PH$ and $4\PH$ vertices by killing terms with two derivatives. The argument goes
as before, the number of diagrams in $\PH\PH \to \PH\PH$ remains the same and so does the
$\mrS\,$-matrix. To be precise there are $s,t,u\,$-channel diagrams and a contact interaction; one 
subtlety is the following: the Lagrangian contains a term linear in $\PH$ whose coefficient 
$\beta_{\PH}$, is fixed, order-by-order, by the request of cancelling $\PH$ 
tadpoles~\cite{Actis:2006ra}. After \eqn{cubic} this term will contribute to $\PH\PH \to \PH\PH$,
with no effect at LO since $\beta_{\PH}$ starts at $\mcO(g^2)$ but with sizable complications 
starting at NLO. Another non negligible complication (even in the unitary gauge) is the presence 
of $19$ new terms induced by the transformation. No ``more transparent'' road here, one gets the
same answer with the same amount of work, not having to jeopardize gauge invariance.
It is worth noting that we have neglected $\Ppz$ and $\Pppm$; to eliminate all terms with two
$\PH$ derivatives in the $\mrR_{\xi}$ gauge we need a cubic transformation mixing the four
fields, $\PH, \Ppz, \Pppm$. Terms with two drivatives of the Higgs-Kibble ghosts will remain,
as expected since the scalar manifold is not flat~\cite{Burgess:2010zq}.

\begin{figure}[ht]
\vspace{-3.cm}
\centering
\includegraphics[width=0.9\textwidth]{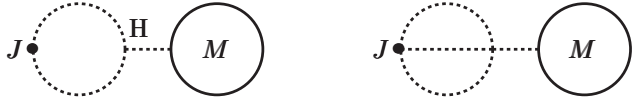}
\vspace{-13.cm}
 \caption{The wave function factors that are requested when using the transformation 
of \eqn{cubic}.}
\label{FTF10}
\end{figure}

\begin{figure}[ht]
\vspace{-3.cm}
\centering
\includegraphics[width=0.9\textwidth]{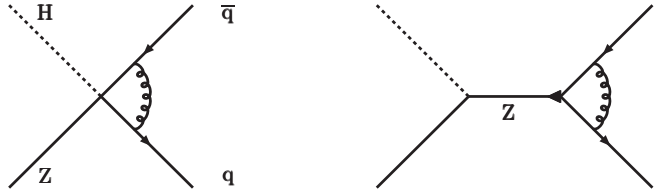}
\vspace{-12.cm}
 \caption{The Phoenix: resurgent contact diagram in $\PH \to \PZ\PAQq\PQq$, even including QCD
corrections. The second diagram is equal to the first.}
\label{FTF11}
\end{figure}

The message to be taken is clear: if special ($\blacktriangleleft$) vertices are not included
the result is wrong since not all of them cancel at the $\mrS\,$-matrix level, as shown in
Fig.~\ref{FTF7}; if they are included then the LO computation of a process is not 
``more transparent'' and becomes considerably involuted at NLO. As an example consider the 
contribution to the $\mrZ\,$-factor for external $\PH$ lines due to the diagrams of 
Fig.~\ref{FTF10}; they obviously cancel against similar diagrams involving $\blacktriangleright$ 
vertices, as long as one is aware of their existence, something that cannot be read from the 
Lagrangian alone without knowing the transformation.
The possibility of ``absorbing'' terms is not limited to $\PH$ tansformations. For instance,
guided by EoM, we could transform $\PZ$ in order to ``absorb'' $\PH\PZ\PAQb\PQb$. This is
achieved by using
\bq
\PZ^{\mu} \to \PZ^{\mu} + i\,g\,g_6\,\frac{\ctw}{2\,M^2}\,\Bigl[
\bigl( \apqo - \apqt \bigr)\,\PAQb\,\gamma^{\mu}\,\gamma_{+}\,\PQb +
\apd\,\PAQb\,\gamma^{\mu}\,\gamma_{-}\,\PQb \Bigr] \spp
\label{Ztrans}
\eq
After cancelling the $\PH\PZ\PAQb\PQb$ terms the transformation has produced the following
additional terms:
\bqa
\Lag_{\add} &=&
   - \frac{i}{2}\,\frac{g\,g_6}{M^2}\,\ctw\,(\apqo - \apqt)\,
         \PAQb\,\gamma^{\mu}\,\gamma_{+}\,\PQb\,(\Box - M^2_0)\,\PZ_{\mu}
\nl
{}&-& \frac{i}{2}\,\frac{g\,g_6}{M^2}\,\ctw\,\apd\,
         \PAQb\,\gamma^{\mu}\,\gamma_{-}\,\PQb\,(\Box - M^2_0)\,\PZ_{\mu}
\nl
{}&-& \frac{1}{24}\,\frac{g^2\,g_6}{M^2}\,(1 + 2\,\ctws)\,\Bigl[
     (\apqo - \apqt)\,
     \PAQb\,\gamma^{\mu}\,\gamma_{+}\,\PQb\,\PAQb\,\gamma_{\mu}\,\gamma_{+}\,\PQb +
     \apd\,
     \PAQb\,\gamma^{\mu}\,\gamma_{+}\,\PQb\,\PAQb\,\gamma_{\mu}\,\gamma_{-}\,\PQb
     \Bigr]
\nl
{}&+& \frac{1}{12}\,\frac{g^2\,g_6}{M^2}\,\stws\,\Bigl[
     (\apqo - \apqt)\,
     \PAQb\,\gamma^{\mu}\,\gamma_{-}\,\PQb\,\PAQb\,\gamma^{\mu}\,\gamma_{+}\,\PQb +
      \apd\,
      \PAQb\,\gamma^{\mu}\,\gamma_{-}\,\PQb\,\PAQb\,\gamma_{\mu}\,\gamma_{-}\,\PQb
      \Bigr] \spc
\eqa
where $\gamma_{\pm} = 1 \pm \gamma^5$ and $M_0 = M/\ctw$. 
\begin{remark}
Special vertices ($\blacktriangleleft$) have been generated but they do not always cancel, only 
when referring to external $\PZ$ lines (\eg in the $\PZ \to \PAQb\PQb$ decay). 
Indeed, they are needed since \eqn{Ztrans} has generated four-fermion interaction terms and there 
is no contribution to the scattering $\PAf\Pf \to \PAf\Pf$ from $\Lag_{\add}$, as it should be.
Furthermore, also the decay $\PH \to \PZ\PAQb\PQb$ remains unchanged. Once again, cancellations 
are there as long as {\underline{all terms are kept}} and no evident advantage is seen at LO 
with increasing complexity at NLO, even in QCD; loop corrections to the contact $\PH\PZ\PAQb\PQb$ 
diagram become loop corrections to a diagram with a special vertex and there is a substantial 
complication with respect to the treatment of $\mrZ$ factors, now the $\PZ$ source can emit/absorb 
a $\PAQb\PQb$ pair. The resurgent contact diagram is depicted in Fig.~\ref{FTF11}; the
$\blacktriangleleft$ in the second diagram, acting on the internal $\PZ$ line, shrinks it to a 
point and reproduces the first diagram (including QCD corrections): transformation does not
kill the Phoenix.
\end{remark}
Proving off-shell WST identities is, of course, another story.
Even this scenario is not complete since we have taken only a subset of terms, essentially those
involving $\PH, \PZ$ and $\PAQb\PQb$. The situation is more involved since we have all the other
terms involving a $\PZ$, among them the $\PZ$ coupling to Faddeev-Popov ghosts ($\mrR_{\xi}$
gauge).
 \clearpage
\bibliographystyle{elsarticle-num}
\bibliography{EFpost}

\begin{thebibliography}{10}
\expandafter\ifx\csname url\endcsname\relax
  \def\url#1{\texttt{#1}}\fi
\expandafter\ifx\csname urlprefix\endcsname\relax\def\urlprefix{URL }\fi
\expandafter\ifx\csname href\endcsname\relax
  \def\href#1#2{#2} \def\path#1{#1}\fi

\bibitem{Costello2011}
K.~Costello, {{Renormalization and Effective Field Theory}}, {Mathematical
  Surveys and Monographs Volume $170$, American Mathematical Society} (2011).

\bibitem{Preskill:1990fr}
J.~Preskill, {Gauge anomalies in an effective field theory}, Annals Phys. 210
  (1991) 323--379.
\newblock \href {http://dx.doi.org/10.1016/0003-4916(91)90046-B}
  {\path{doi:10.1016/0003-4916(91)90046-B}}.

\bibitem{Einhorn:2013kja}
M.~B. Einhorn, J.~Wudka, {The Bases of Effective Field Theories}, Nucl. Phys.
  B876 (2013) 556--574.
\newblock \href {http://arxiv.org/abs/1307.0478} {\path{arXiv:1307.0478}},
  \href {http://dx.doi.org/10.1016/j.nuclphysb.2013.08.023}
  {\path{doi:10.1016/j.nuclphysb.2013.08.023}}.

\bibitem{YR4}
D.~de~Florian, et~al., {Handbook of LHC Higgs Cross Sections: 4. Deciphering
  the Nature of the Higgs Sector}\href {http://arxiv.org/abs/1610.07922}
  {\path{arXiv:1610.07922}}.

\bibitem{NLOnote}
G.~Passarino, M.~Trott, {{The Standard Model Effective Field Theory and Next to
  Leading Order}}, {LHCHXSWG-DRAFT-INT-2016-005,
  https://cds.cern.ch/record/$2138031$} (2016).

\bibitem{Passarino:2016pzb}
G.~Passarino, M.~Trott,
  \href{http://inspirehep.net/record/1494607/files/arXiv:1610.08356.pdf}{{The
  Standard Model Effective Field Theory and Next to Leading Order}}, 2016.
\newblock \href {http://arxiv.org/abs/1610.08356} {\path{arXiv:1610.08356}}.
\newline\urlprefix\url{http://inspirehep.net/record/1494607/files/arXiv:1610.0%
8356.pdf}

\bibitem{Lehmann:1957zz}
H.~Lehmann, K.~Symanzik, W.~Zimmermann, {On the formulation of quantized field
  theories. II}, Nuovo Cim. 6 (1957) 319--333.
\newblock \href {http://dx.doi.org/10.1007/BF02832508}
  {\path{doi:10.1007/BF02832508}}.

\bibitem{'tHooft:1972ue}
G.~'t~Hooft, M.~J.~G. Veltman, {Combinatorics of gauge fields}, Nucl. Phys. B50
  (1972) 318--353.
\newblock \href {http://dx.doi.org/10.1016/S0550-3213(72)80021-X}
  {\path{doi:10.1016/S0550-3213(72)80021-X}}.

\bibitem{Exa}
G.~t~Hooft, M.~J.~G. Veltman, {{Example of a gauge field theory}}, {Proceedings
  of the Colloquium on Renormalization of Yang-Mills Fields, Marseille, June
  19-23, 1972.} (1972).

\bibitem{'tHooft:1973pz}
G.~'t~Hooft, M.~J.~G. Veltman, {DIAGRAMMAR}, NATO Sci. Ser. B 4 (1974)
  177--322.

\bibitem{Chisholm:1961tha}
J.~S.~R. Chisholm, {Change of variables in quantum field theories}, Nucl. Phys.
  26~(3) (1961) 469--479.
\newblock \href {http://dx.doi.org/10.1016/0029-5582(61)90106-7}
  {\path{doi:10.1016/0029-5582(61)90106-7}}.

\bibitem{Kamefuchi:1961sb}
S.~Kamefuchi, L.~O'Raifeartaigh, A.~Salam, {Change of variables and equivalence
  theorems in quantum field theories}, Nucl. Phys. 28 (1961) 529--549.

\bibitem{Kallosh:1972ap}
R.~E. Kallosh, I.~V. Tyutin, {The Equivalence theorem and gauge invariance in
  renormalizable theories}, Yad. Fiz. 17 (1973) 190--209, [Sov. J. Nucl.
  Phys.17,98(1973)].

\bibitem{Burgess:2010zq}
C.~P. Burgess, H.~M. Lee, M.~Trott, {Comment on Higgs Inflation and
  Naturalness}, JHEP 07 (2010) 007.
\newblock \href {http://arxiv.org/abs/1002.2730} {\path{arXiv:1002.2730}},
  \href {http://dx.doi.org/10.1007/JHEP07(2010)007}
  {\path{doi:10.1007/JHEP07(2010)007}}.

\bibitem{Arzt:1993gz}
C.~Arzt, {Reduced effective Lagrangians}, Phys. Lett. B342 (1995) 189--195.
\newblock \href {http://arxiv.org/abs/hep-ph/9304230}
  {\path{arXiv:hep-ph/9304230}}, \href
  {http://dx.doi.org/10.1016/0370-2693(94)01419-D}
  {\path{doi:10.1016/0370-2693(94)01419-D}}.

\bibitem{Grzadkowski:2010es}
B.~Grzadkowski, M.~Iskrzynski, M.~Misiak, J.~Rosiek, {Dimension-Six Terms in
  the Standard Model Lagrangian}, JHEP 1010 (2010) 085.
\newblock \href {http://arxiv.org/abs/1008.4884} {\path{arXiv:1008.4884}},
  \href {http://dx.doi.org/10.1007/JHEP10(2010)085}
  {\path{doi:10.1007/JHEP10(2010)085}}.

\bibitem{Wudka:1994ny}
J.~Wudka, {Electroweak effective Lagrangians}, Int.J.Mod.Phys. A9 (1994)
  2301--2362.
\newblock \href {http://arxiv.org/abs/hep-ph/9406205}
  {\path{arXiv:hep-ph/9406205}}, \href
  {http://dx.doi.org/10.1142/S0217751X94000959}
  {\path{doi:10.1142/S0217751X94000959}}.

\bibitem{Gervais:1976ws}
J.-L. Gervais, A.~Jevicki, {Point Canonical Transformations in Path Integral},
  Nucl. Phys. B110 (1976) 93--112.
\newblock \href {http://dx.doi.org/10.1016/0550-3213(76)90422-3}
  {\path{doi:10.1016/0550-3213(76)90422-3}}.

\bibitem{Salomonson:1976yn}
P.~Salomonson, {When Does a Nonlinear Point Transformation Generate an Extra
  $\mcO(\hbar^2)$ Potential in the Effective Lagrangian?}, Nucl. Phys. B121
  (1977) 433--444.
\newblock \href {http://dx.doi.org/10.1016/0550-3213(77)90165-1}
  {\path{doi:10.1016/0550-3213(77)90165-1}}.

\bibitem{Vilkovisky:1984st}
G.~A. Vilkovisky, {The Unique Effective Action in Quantum Field Theory}, Nucl.
  Phys. B234 (1984) 125--137.
\newblock \href {http://dx.doi.org/10.1016/0550-3213(84)90228-1}
  {\path{doi:10.1016/0550-3213(84)90228-1}}.

\bibitem{Batalin:2012rv}
I.~A. Batalin, K.~Bering, {Reparametrization-Invariant Effective Action in
  Field-Antifield Formalism}, Int. J. Mod. Phys. A28 (2013) 1350027.
\newblock \href {http://arxiv.org/abs/1211.6391} {\path{arXiv:1211.6391}},
  \href {http://dx.doi.org/10.1142/S0217751X13500279}
  {\path{doi:10.1142/S0217751X13500279}}.

\bibitem{Kreimer:2016jxo}
D.~Kreimer, K.~Yeats, {Diffeomorphisms of quantum fields}\href
  {http://arxiv.org/abs/1610.01837} {\path{arXiv:1610.01837}}.

\bibitem{Finkemeier:1997re}
M.~Finkemeier, H.~Georgi, M.~McIrvin, {Reparametrization invariance revisited},
  Phys. Rev. D55 (1997) 6933--6943.
\newblock \href {http://arxiv.org/abs/hep-ph/9701243}
  {\path{arXiv:hep-ph/9701243}}, \href
  {http://dx.doi.org/10.1103/PhysRevD.55.6933}
  {\path{doi:10.1103/PhysRevD.55.6933}}.

\bibitem{Luke:1992cs}
M.~E. Luke, A.~V. Manohar, {Reparametrization invariance constraints on heavy
  particle effective field theories}, Phys. Lett. B286 (1992) 348--354.
\newblock \href {http://arxiv.org/abs/hep-ph/9205228}
  {\path{arXiv:hep-ph/9205228}}, \href
  {http://dx.doi.org/10.1016/0370-2693(92)91786-9}
  {\path{doi:10.1016/0370-2693(92)91786-9}}.

\bibitem{Passarino:2012cb}
G.~Passarino, {NLO Inspired Effective Lagrangians for Higgs Physics}, Nucl.
  Phys. B868 (2013) 416--458.
\newblock \href {http://arxiv.org/abs/1209.5538} {\path{arXiv:1209.5538}},
  \href {http://dx.doi.org/10.1016/j.nuclphysb.2012.11.018}
  {\path{doi:10.1016/j.nuclphysb.2012.11.018}}.

\bibitem{Ghezzi:2015vva}
M.~Ghezzi, R.~Gomez-Ambrosio, G.~Passarino, S.~Uccirati, {NLO Higgs effective
  field theory and kappa-framework}, JHEP 07 (2015) 175.
\newblock \href {http://arxiv.org/abs/1505.03706} {\path{arXiv:1505.03706}},
  \href {http://dx.doi.org/10.1007/JHEP07(2015)175}
  {\path{doi:10.1007/JHEP07(2015)175}}.

\bibitem{Actis:2006ra}
S.~Actis, A.~Ferroglia, M.~Passera, G.~Passarino, {Two-Loop Renormalization in
  the Standard Model. Part I: Prolegomena}, Nucl.Phys. B777 (2007) 1--34.
\newblock \href {http://arxiv.org/abs/hep-ph/0612122}
  {\path{arXiv:hep-ph/0612122}}, \href
  {http://dx.doi.org/10.1016/j.nuclphysb.2007.04.021}
  {\path{doi:10.1016/j.nuclphysb.2007.04.021}}.

\end{thebibliography}

\end{document}